\documentclass[aps,prc,groupedaddress,showpacs,manuscript]{revtex4}

\usepackage{graphicx}

\begin{document}

\title{Probing the equation of state with pions}

\author {Qingfeng Li$\, ^{1}$\footnote{Fellow of the Alexander von Humboldt Foundation.}
\email[]{Qi.Li@fias.uni-frankfurt.de}, Zhuxia Li$\, ^{2}$
\email[]{lizwux@iris.ciae.ac.cn}, Sven Soff$\, ^{3}$, Marcus
Bleicher$\, ^{3}$, and Horst St\"{o}cker$\, ^{1,3}$}
\address{
1) Frankfurt Institute for Advanced Studies (FIAS), Johann Wolfgang Goethe-Universit\"{a}t, Max-von-Laue-Str.\ 1, D-60438 Frankfurt am Main, Germany\\
2) China Institute of Atomic Energy, P.O.\ Box 275 (18),
Beijing 102413, P.R.\ China\\
3) Institut f\"{u}r Theoretische Physik, Johann Wolfgang Goethe-Universit\"{a}t, Max-von-Laue-Str.\ 1, D-60438 Frankfurt am Main, Germany\\
 }


\begin{abstract}
The influence of the isospin-independent, isospin- and
momentum-dependent equation of state (EoS), as well as the Coulomb
interaction on the pion production in intermediate energy heavy
ion collisions (HICs) is studied for both isospin-symmetric and
neutron-rich systems. The Coulomb interaction plays an important
role in the reaction dynamics, and strongly influences the
rapidity and transverse momentum distributions of charged pions.
It even leads to the $\pi^-/\pi^+$ ratio deviating slightly from
unity for isospin-{\it symmetric} systems. The Coulomb interaction
between mesons and baryons is also crucial for reproducing the
proper pion flow since it changes the behavior of the directed and
the elliptic flow components of pions visibly.

The EoS can be better investigated in neutron-rich system if {\it
multiple} probes are measured simultaneously. For example, the
rapidity and the transverse momentum distributions of the charged
pions, the $\pi^-/\pi^+$ ratio, the various pion flow components,
as well as the {\it difference} of $\pi^+$-$\pi^-$ flows. A new
sensitive observable is proposed to probe the symmetry potential energy
at high densities, namely the transverse momentum distribution of
the elliptic flow difference [$\Delta v_2^{\pi^+ -
\pi^-}(p_t^{\rm c.m.})$].
\end{abstract}


\pacs{24.10.Lx, 25.75.Dw, 25.75.-q} \maketitle

\section{Introduction}
Intermediate energy heavy ion collisions (HICs) are closely
connected to the investigation of the nuclear equation of state
(EoS). One of the main issues is to pin down the incompressibility
($K_{\rm NM}$) of nuclear matter.  Although many efforts have been
pursued in the past few decades this problem is still far from
being solved thoroughly (see, for example, Refs.\
\cite{Vre03,Riz04}). Furthermore, the symmetry energy
\cite{Tsa01,Fur02,Che04,Kho05} for isospin-asymmetric matter, the
momentum dependence \cite{Ai87,Gale90,Par95,Bass95,Gre99,LiB05} of
the EoS, have also been found to be very important for the
dynamics of the intermediate energy HICs and make the problem of
probing the EoS even more complex.

In order to explore explicitly the incompressibility, the isospin
dependence and the momentum dependence of the EoS, sensitive
probes were put forward, however, mostly individually. Nevertheless, sensitive probes are not always proprietary, that is, they
might be affected not only by single physical quantity. Therefore,
it is quite necessary to explore multiple probes simultaneously so
that the comparison between the experimental data and the
corresponding theoretical predictions becomes more
consistent.

In addition to free nucleons and light fragments, probes related
to pions such as the $\pi^{-}/\pi^{+}$ ratio and the directed and
elliptic flows, have been proven to be very useful to test
the reaction dynamics as well as the EoS
\cite{Bass95,Mah98,Wag00,LiB05,LiQ05,LiQ052,Dan00,Dan002,LiB96,LiB99,Bra00}.
It was supposed and observed in Refs. \cite{Bass95,LiB96,Kin97} that pions show a weak
positive flow effect in central collisions while a weak antiflow
effect in peripheral collisions due to the shadowing effect of
spectators. It should be mentioned that experimental measurements
on the pion production with various collision systems and beam
energies have been released by the FOPI/GSI Collaboration in
recent years, for example, see Refs. \cite{Pel97,Hong05}, which
indeed largely deepen our insight into the production mechanism of
pions in intermediate energy HICs, as well as the dynamics of the
nuclear reaction itself.

In this work we will study a variety of observables related to
pion production in isospin-symmetric and -asymmetric systems from
central and semi-peripheral intermediate-energy HICs. The
microscopic transport model - ultrarelativistic quantum molecular
dynamics (UrQMD) \cite{Bass98,Bleicher99,Reiter03,Bra04} - is
adopted with an update of the potentials in the mean field part.

The paper is arranged as follows. In section II, we briefly
introduce the UrQMD model and the improvements of the potentials
in the mean field part. In section III, the results of the ratios
between the yields of charged pions, the various directed and
elliptic flows of pion mesons are shown and discussed. Finally, a
summary is given in section IV.

\section{The UrQMD transport model and the potential update}
In the UrQMD transport model \cite{Bass98,Bleicher99}, the
initialization of projectile and target nuclei, the equation of
motion of hadrons, and the collision term are described
microscopically. This model was designed at the beginning for
simulating HICs in the energy range from SIS to RHIC, where the
contribution of nuclear mean field potentials to the dynamics of
the reaction is considered to be weak. It is known that the
conventional (or isospin-dependent) quantum molecular dynamics
(QMD) model \cite{Bass95,Aic91} is mainly applied to the
(isospin-asymmetric) intermediate-energy HICs. The UrQMD model
inherits the basic treatment of the baryonic equation of motion in
the QMD model, thus, after introducing some modern ingredients for
the mean field part, it is believed that the UrQMD model can also
be used to properly describe the physical phenomena in HICs at
intermediate energies.

Furthermore, it has been seen that the contribution of the mean
field to the reaction dynamics can not simply be neglected at SIS
and even AGS energies \cite{LiQ052,Soff99}. It was shown in Ref.
\cite{Soff99} that the experimental elliptic flow as a function of
beam energy ranging from $2-10A$ GeV for midrapidity protons in
Au+Au collisions can be only successfully described if mean field
potentials are taken into account, while pure cascade calculations
fail. In Ref. \cite{LiQ052} it has been shown that the potential
that $\Sigma$ hyperons encounter in the nuclear medium plays an
important role in their evolution process in HICs at SIS energies.
Thus, we have updated the potential interactions in the UrQMD model for the
intended studies.

In addition to the isospin- and momentum- independent terms
originally implemented in the mean field part of the UrQMD model,
the following contributions are also considered here:

1. The contribution of the Coulomb interaction between {\it mesons} and hadrons
(mesonic Coulomb potential) are supplied in addition to the
Coulomb interaction between two baryons (baryonic Coulomb potential),
in total, we call it hadronic Coulomb potential. It has been found that
the Coulomb interaction between baryons and charged $\pi$ mesons
plays an essential role in the dependence of the $\pi^-/\pi^+$
ratios on rapidity and transverse momentum
\cite{Pel97,Hong05,LiQ05}.

\begin{figure}
\includegraphics[angle=0,width=0.8\textwidth]{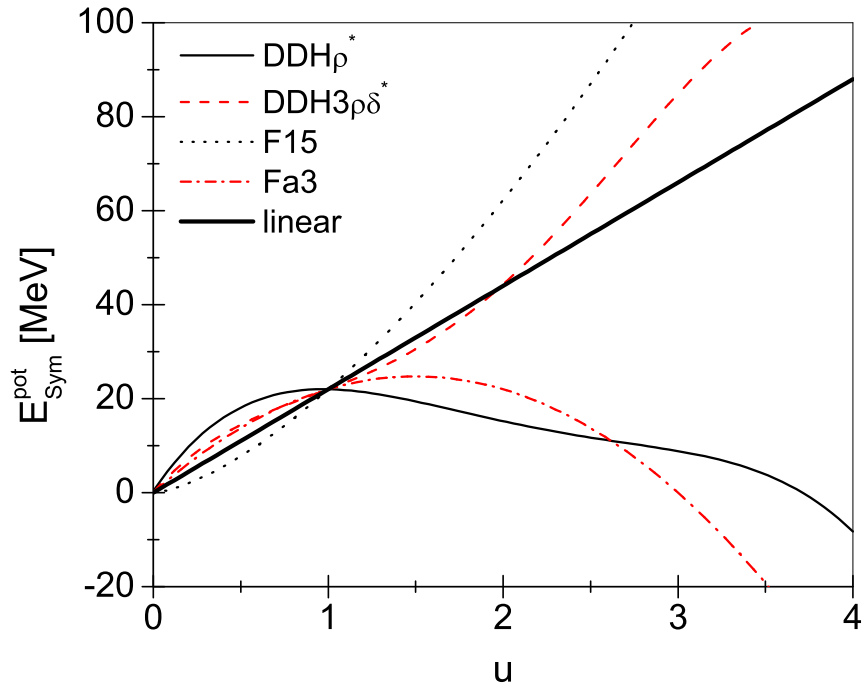}
\caption{Parametrizations of the nuclear symmetry potential energy
DDH$\rho^*$, DDH3$\rho\delta^*$, F15, and Fa3, as well as a linear
one as a function of the reduced density $u$.} \label{fig1}
\end{figure}

2. Like in our previous work \cite{LiQ05,LiQ052}, the symmetry
potentials of all baryons, i.e., the nucleons, the
$\Delta(1232)$s, the $N^*(1440)$s, and the hyperons $\Lambda$ and
$\Sigma$, are introduced. Four density-dependent parameterizations
for symmetry potential energy are considered: (1) $u^\gamma$ with
$\gamma=1.5$ (called F15). Here $u=\rho/\rho_0$ is the reduced
nuclear density; (2) $u\cdot(a-u)/(a-1)$ with $a=3$ (Fa3). $a$ is
the so-called reduced critical density \cite{LiB02}; (3) and (4) so-called
DDH$\rho^*$ and DDH3$\rho\delta^*$ symmetry potential energies, which are
inspired by the relativistic mean-field calculations of DDH$\rho$
and DDH3$\rho\delta$ \cite{Gai04}. The symmetry energy coefficient
$S_0=34$ MeV is adopted \cite{Vre03,Dal04}. The density dependence
of the symmetry potential energies adopted is shown in Fig.\
\ref{fig1}. One sees from Fig.\ \ref{fig1} that for reduced
densities $u<1$, DDH3$\rho\delta^*$ is very close to Fa3, both lie
between DDH$\rho^*$ and F15. For $u>1$, their density dependences
are rather different: for $1<u<2.6$, the order of the symmetry
potential energy $E_{\rm sym}^{\rm pot}$ is F15 $>$
DDH3$\rho\delta^*$ $>$ Fa3 $>$ DDH$\rho^*$, while for $u>2.6$, F15
$>$ DDH3$\rho\delta^*$ $>$ DDH$\rho^*$ $>$ Fa3.

3. Momentum-dependent interactions for all baryons are
introduced. The form of the momentum dependence is taken from the
IQMD model \cite{Aic87,Bass95}, which reads

\begin{equation}
U_{\rm md}=t_{\rm md} \ln^2[1+a_{\rm md} (\Delta
{\bf p})^2]u, \label{umd}
\end{equation}
in which $\Delta
{\bf p}={\bf p}_i-{\bf p}_j$
represents the relative momentum of two nucleons $i$ and $j$. The
parametrizations of $t_{\rm md}$ and $a_{\rm md}$ are listed in
Table \ref{tab1}. We note that B.-A. Li et al. have used an
isospin-dependent momentum-dependent parametrization in the BUU
model, which is guided by a Hartree-Fock calculation using the
Gogny effective interaction \cite{LiB04,LiB042}. However, here we
do not consider the isospin dependence in the momentum dependent
part of the mean field.

Besides the two- and three-body Skyrme potential ($U_{\rm
sky}=\alpha\, u + \beta\, u^{\gamma_{\rm sky}}$), similar to the
IQMD model \cite{Bass95,Soff95}, we consider a variant nuclear
incompressibility, namely, a hard EoS with $K=300$ MeV ("H-EoS"),
a soft EoS with $K=200$ MeV ("S-EoS"), a hard EoS ($K=380$ MeV)
with momentum dependence ("HM-EoS"), and a soft EoS ($K=200$ MeV)
with momentum dependence ("SM-EoS"). The parameters of the various
EoS used in the UrQMD model are listed in Table \ref{tab1}. In
order to reproduce the ground-state properties of finite nuclei in
the UrQMD model (for instance, the binding energy $E_B$ and the
r.m.s radius), the parameters have been slightly readjusted. Note, that the potentials used here are not included in the currently available version of the model, but will be included in future versions.

\begin{table}

\caption{Parameter sets for the nuclear equation of state used in
the extended version of the UrQMD model.}

\begin{tabular}{l|ccc|ccc}
\hline\hline
EoS& $\alpha$\, [MeV] & $\beta$\, [MeV] & $\gamma_{\rm sky}$ & $t_{\rm md}$ [MeV] & $a_{\rm md}$ [$\frac{c^2}{\rm GeV^2}$] & $S_0$ [MeV]  \\
\hline
H & -165 & 126&1.676 & - & - & 34\\
S & -353 & 304&7/6 & - & - & 34\\
HM & -138 & 60&2.08 & 1.57 & 500 & 34\\
SM & -393 & 320&1.14 & 1.57 & 500 & 34\\
\hline\hline

\end{tabular}
\label{tab1}
\end{table}

\section{Results and Discussions}
\begin{figure}
\includegraphics[angle=0,width=0.8\textwidth]{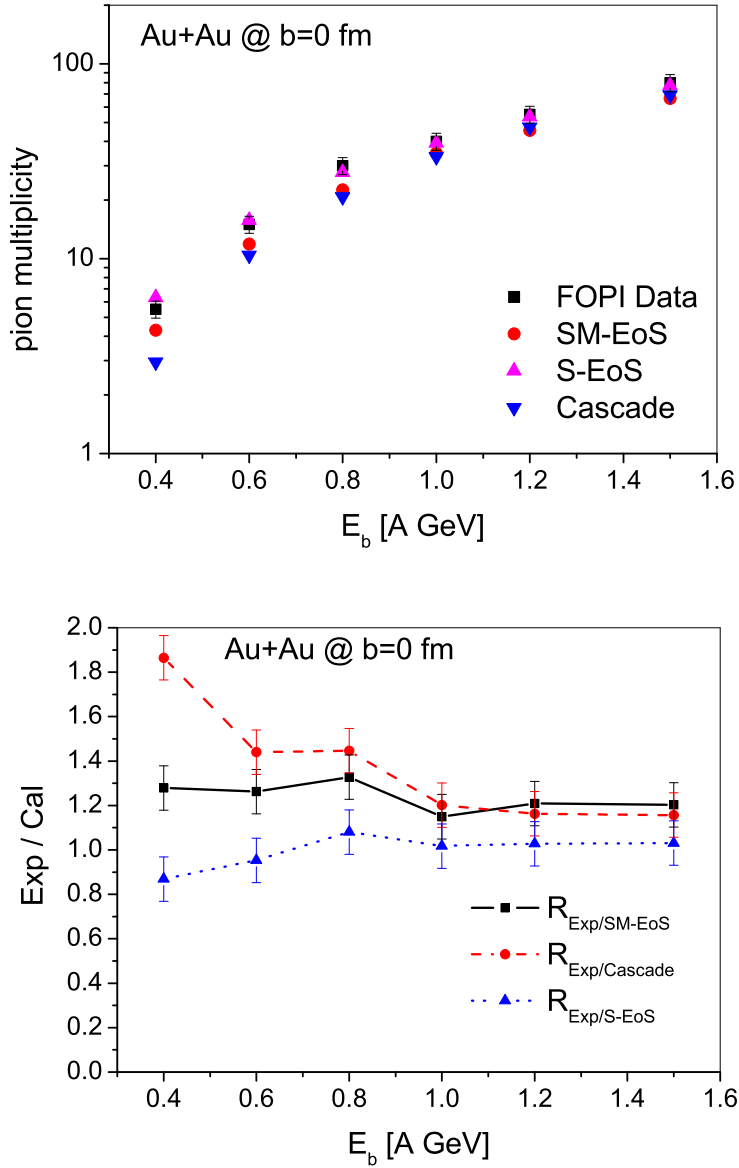}
\caption{Upper plot: The excitation function of the pion
multiplicities for central ($b=0\ {\rm fm}$) $^{197}$Au+$^{197}$Au
reaction. The Cascade-mode, S-EoS, and SM-EoS are adopted for
calculations. The FOPI preliminary data for central Au+Au collisions are also shown
\cite{FOPI05}. Lower plot: The ratios between the experimental
data and the calculations.} \label{fig2}
\end{figure}

Firstly, we test the applicability of the UrQMD model to
intermediate energy HICs. Fig.\ \ref{fig2} shows the results for
the excitation function of pion multiplicities in central ($b=0\
{\rm fm}$) $^{197}$Au+$^{197}$Au reactions at energies from $0.4A$
to $1.5A\ {\rm GeV}$ and the comparison with the recent FOPI
preliminarily experimental data \cite{FOPI05}. The ratios
between the calculations (with S- and SM-EoS and without
potentials, namely the cascade mode) and the experimental data are
shown in the lower plot. In Ref.\,\cite{Kol04} the multiplicities
of pions were compared within various transport approaches for
HICs around $1A\ {\rm GeV}$. A significant variation of the total
pion yield with the different approaches has been found.
Generally, these transport models overpredict the pion yields.
Overall, our calculations (Fig.\ \ref{fig2} with EoS, especially
with a S-EoS) are in good agreement with data, while the results
without mean-field potential (the cascade mode) deviate from the
experimental data at lower beam energies ($<0.8A\ {\rm GeV}$).

\begin{figure}
\includegraphics[angle=0,width=0.8\textwidth]{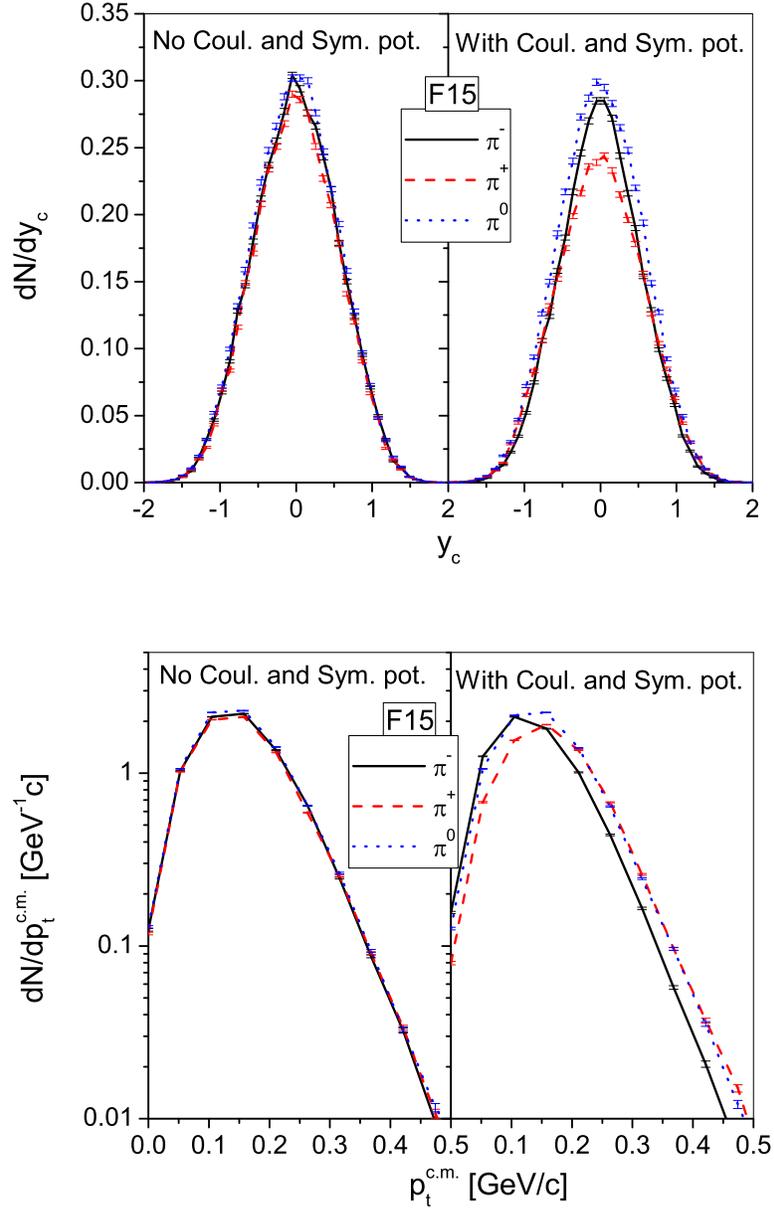}
\caption{The rapidity (upper plots) and the transverse momentum
(lower plots) distributions of pions with (right plots) or without
(left plots) hadronic Coulomb and baryonic symmetry potentials.
The isospin-symmetric reaction $^{40}{\rm Ca}+^{40}{\rm Ca}$ at
the beam energy $E_b=0.4A$ GeV with the impact parameter $b=0$ fm
is chosen. } \label{fig3}
\end{figure}

Now let us turn to see how both the Coulomb and the symmetry
potentials influence the production of pions in isospin-symmetric
HICs.  Fig.\ \ref{fig3} shows the rapidity (upper plots) and the
transverse momentum (lower plots) distributions (in the
center-of-mass system) of pions in central ($b=0$ fm) collisions
$^{40}{\rm Ca}+^{40}{\rm Ca}$ at beam energy $E_b=0.4A$ GeV. The
SM-EoS and the F15 symmetry potential energy are selected here. In the
left (right) two plots we show the distributions without (with)
the contribution of both Coulomb (of all charged hadrons) and
symmetry potentials. When the Coulomb and symmetry potentials are
switched off, the multiplicities of $\pi^-$, $\pi^+$, and $\pi^0$
are equal to each other within statistical errors, which means that the
isospin symmetry is preserved, as one expects. While the Coulomb
and symmetry potentials are switched on, the rapidity and
transverse momentum distributions of $\pi^-$, $\pi^+$, and $\pi^0$
mesons are clearly different. At mid-rapidity and in the low
transverse momentum region, the multiplicity of $\pi^+$ mesons is
smaller than the $\pi^-$ ones, and vice versa in the region of the
projectile-target rapidity or large transverse momentum. We notice
that, after integrating the multiplicities of $\pi^-$ and $\pi^+$
mesons separately, the $\pi^-/\pi^+$ ratio is about $1.05$. We
have also checked this ratio at a higher beam energy, $E_{\rm
b}=0.8A$ GeV, and found that it is reduced to $1.02$.

\begin{figure}
\includegraphics[angle=0,width=0.8\textwidth]{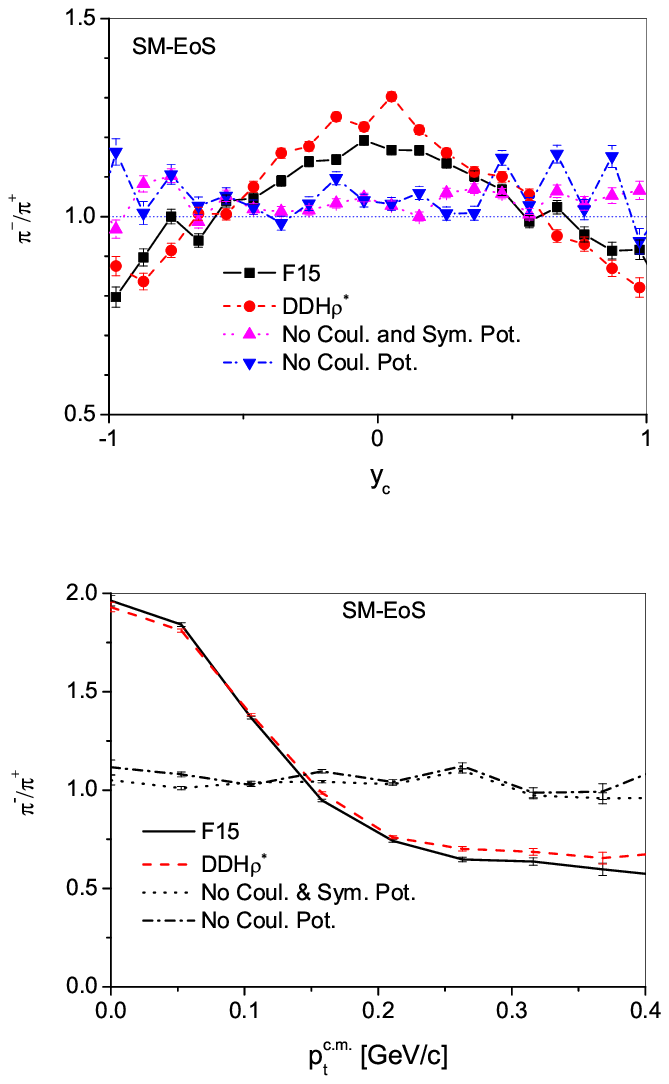}
\caption{The rapidity (upper plot) and the transverse momentum
(lower plot) distributions of the $\pi^-/\pi^+$ ratio as
calculated (1) without symmetry and the Coulomb potentials ("No
Coul.\ and Sym.\ Pot."); (2) with symmetry and Coulomb potentials
(cases "F15" and "DDH$\rho^*$"); and (3) without Coulomb potential
but with symmetry potential DDH$\rho^*$ ("No Coul.\ Pot."). The
reaction $^{40}{\rm Ca}+^{40}{\rm Ca}$ at $E_b=0.4A$ GeV with
$b=0$ fm is chosen.} \label{fig4}
\end{figure}

Fig.\ \ref{fig4} shows the rapidity (upper plot) and the
transverse momentum (lower plot) distributions of the
$\pi^{-}/\pi^{+}$ ratio for the same reaction as in Fig.\
\ref{fig3}. In the plots we show the results for the following
cases (1) without hadronic Coulomb and baryonic symmetry
potentials ("No Coul. and Sym. Pot.") (2) with hadronic Coulomb
and baryonic symmetry potentials F15 and DDH$\rho^*$ ("F15" and
"DDH$\rho^*$") and (3) without the hadronic Coulomb potential but
with the DDH$\rho^*$ symmetry potential energy ("No Coul.\ Pot."). One
finds that without the Coulomb and symmetry potentials the
$\pi^{-}/\pi^{+}$ ratio is around unity. When the Coulomb and
symmetry potentials are switched on, the $\pi^-/\pi^+$ ratio
depends weakly on the symmetry potential. At mid-rapidity and at
large transverse momenta, the softer the symmetry potential is, the larger
the $\pi^-/\pi^+$ ratio is, and the other way around in the
regions of projectile-target rapidity and small transverse
momentum. When the Coulomb interaction is switched off but the
symmetry potential is on, i.e., case (3), the $\pi^-/\pi^+$ ratio
is again around unity. This means that the Coulomb interaction
is the leading cause of the $\pi^-/\pi^+$ ratio deviating from
unity, which is shown in case (2). When the Coulomb potential is
taken into account, the protons and also the positively charged
$\Delta$s are slightly pushed into the lower density region so
that more neutrons and negatively charged $\Delta$s will be in the
high-density region. As a consequence, the isospin asymmetry
$(\rho_{n}-\rho_{p})/(\rho_{n}+\rho_{p})$ will be different from
zero locally although the total system is isospin-symmetric. Thus,
the symmetry potential begins to play a role that leads to the
$\pi^-/\pi^+$ ratio depending on the symmetry potential for case
(2). So we expect that the deviation of the $\pi^-/\pi^+$ ratio
from unity depends on rapidity and transverse momentum and helps
to determine the symmetry potential for isospin-symmetric systems.

\begin{figure}
\includegraphics[angle=0,width=0.8\textwidth]{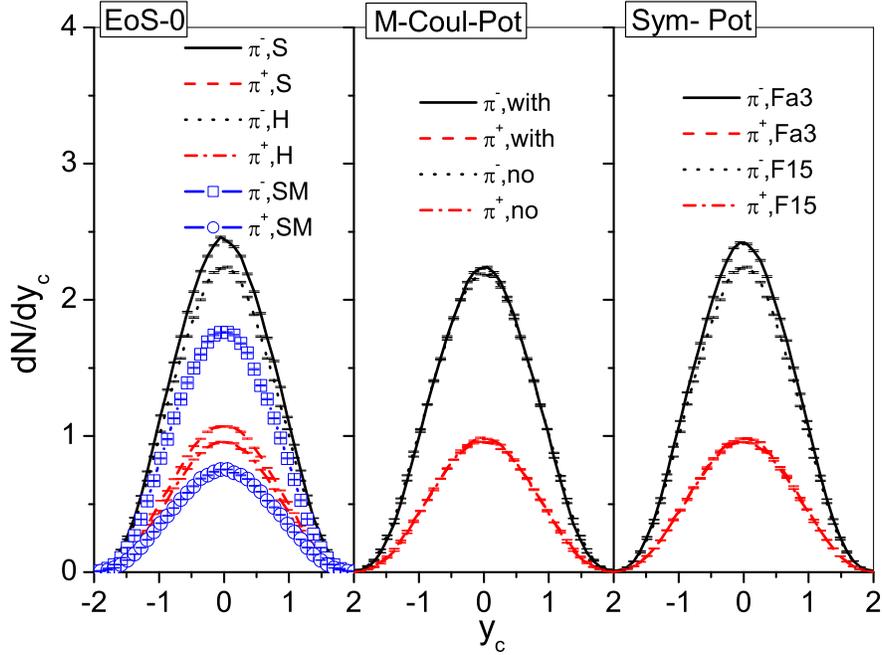}
\caption{Rapidity distributions of $\pi^-$ and $\pi^+$
for various isospin-independent EoS (S-, H,
and SM-EoS, left plot), with or without mesonic Coulomb potential
(middle plot), and with the F15 and Fa3 symmetry potentials (right
plot) (see text). The reaction $^{208}{\rm Pb}+^{208}{\rm Pb}$ at $E_b=0.8A$
GeV and $b=7-9$ fm is chosen.} \label{fig5}
\end{figure}

\begin{figure}
\includegraphics[angle=0,width=0.8\textwidth]{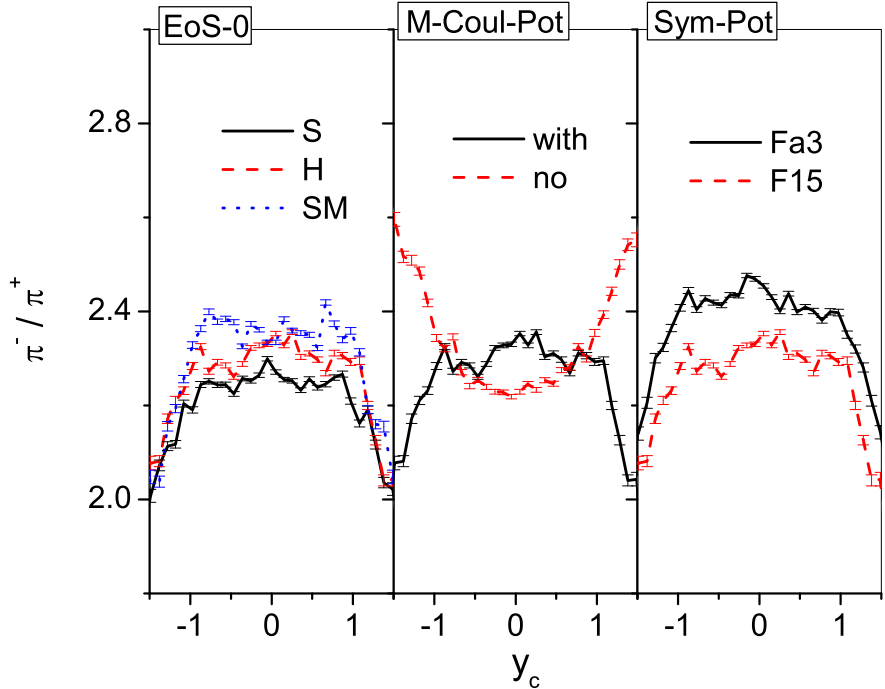}
\caption{Rapidity distributions of $\pi^-/\pi^+$ ratios
for various isospin-independent EoS (S-, H,
and SM-EoS, left plot), with or without mesonic Coulomb potential
(middle plot), and with the F15 and Fa3 symmetry potentials (right
plot) (see text). The reaction $^{208}{\rm Pb}+^{208}{\rm Pb}$ at $E_b=0.8A$
GeV and $b=7-9$ fm is chosen.} \label{fig6}
\end{figure}

Before investigating the influence of various mean-field
potentials on the pion flow in intermediate energy HICs, it is
necessary to show the rapidity distributions of $\pi^-$ and
$\pi^+$ mesons (Fig.\ \ref{fig5}), as well as the $\pi^-/\pi^+$
ratios (Fig.\ \ref{fig6}). They are calculated (1) with different
isospin-independent EoS, that is, the H-, S-, and SM-EoS (left
plot, titled "EoS-0") but with the same symmetry potential F15 and
including the hadronic Coulomb interaction; 2) with or without
mesonic Coulomb potential (middle plot, titled "M-Coul-Pot") but
with the same H-EoS and the symmetry potential F15; and (3) with
different symmetry potentials F15 and Fa3 (right plot, titled
"Sym-Pot") but with the same H-EoS and the hadronic Coulomb
potential. Semi-peripheral ($b=7-9$ fm) $^{208}{\rm Pb}+^{208}{\rm
Pb}$ collisions at $E_b=0.8A$ GeV are studied for all following
calculations. From Fig.\ \ref{fig5} we can see, at first glance,
the uncertainty of the isospin-independent EoS obviously affects
the total pion multiplicity; the effect of the density dependence
of the symmetry potential on the $\pi^-$ multiplicity is visible,
while the mesonic Coulomb potential hardly affects the
multiplicity of charged pions in the neutron-rich $^{208}{\rm
Pb}+^{208}{\rm Pb}$ system. These characteristics have already
been seen in \cite{LiQ05}. The left plot of Fig.\ \ref{fig5} shows
that the pion yield with momentum dependent interactions (SM-EoS)
is obviously lower than that with S-EoS or H-EoS, which has been
shown in Ref.\ \cite{Ai87}. Further, we see that both the mesonic
Coulomb and the symmetry potentials hardly affect the $\pi^+$
multiplicities in neutron-rich HICs. Therefore, the rapidity
distribution of $\pi^+$ can provide more accurate information on
the isospin-independent part of the EoS for the neutron-rich HICs.

The $\pi^-/\pi^+$ ratio at $-1<y_{c}<1$ is smaller for a soft EoS as compared to a
hard EoS (left plot of Fig.\ \ref{fig6}), due to more two-body scatterings.
The $\pi^-/\pi^+$ ratio with a soft symmetry
potential is larger than the one with a hard symmetry potential (right plot).
The effect of the symmetry potential has been studied extensively in
\cite{LiQ05,LiQ052}. Here we want to stress that both, the
isospin-independent and isospin-dependent EoS parts, obviously affect the
$\pi^-/\pi^+$ ratio but in the opposite direction when the
stiffness changes from soft to hard. When the mesonic Coulomb potential is not taken into account (middle plot)
the $\pi^-/\pi^+$ ratio is larger at rapidities exceeding one than at mid-rapidity, which was proven wrong by experimental measurements, for example, see Ref.\ \cite{Hong05}. It
means that the mesonic Coulomb potential needs to be considered, besides the baryonic Coulomb potential. One
further finds from this figure that the $\pi^-/\pi^+$ ratio becomes rather flat at
small rapidities ($-0.75 < y_{c} < 0.75$) in
the semi-peripheral HICs. We have also found a double peaked
structure (in the projectile and target regions) of the rapidity
distribution of the $\pi^-/\pi^+$ ratio at even larger impact
parameters. This behavior is rather different from central
collisions (upper plot of Fig.\ \ref{fig4})
where the $\pi^-/\pi^+$ ratio is peaked in the mid-rapidity region. This
change of the behavior of the $\pi^-/\pi^+$ ratio as a function of
rapidity from central to semi-peripheral HICs is
due to the decrease of the number of nucleon-nucleon
collisions \cite{Ai87,Mah98}.

\begin{figure}
\includegraphics[angle=0,width=0.8\textwidth]{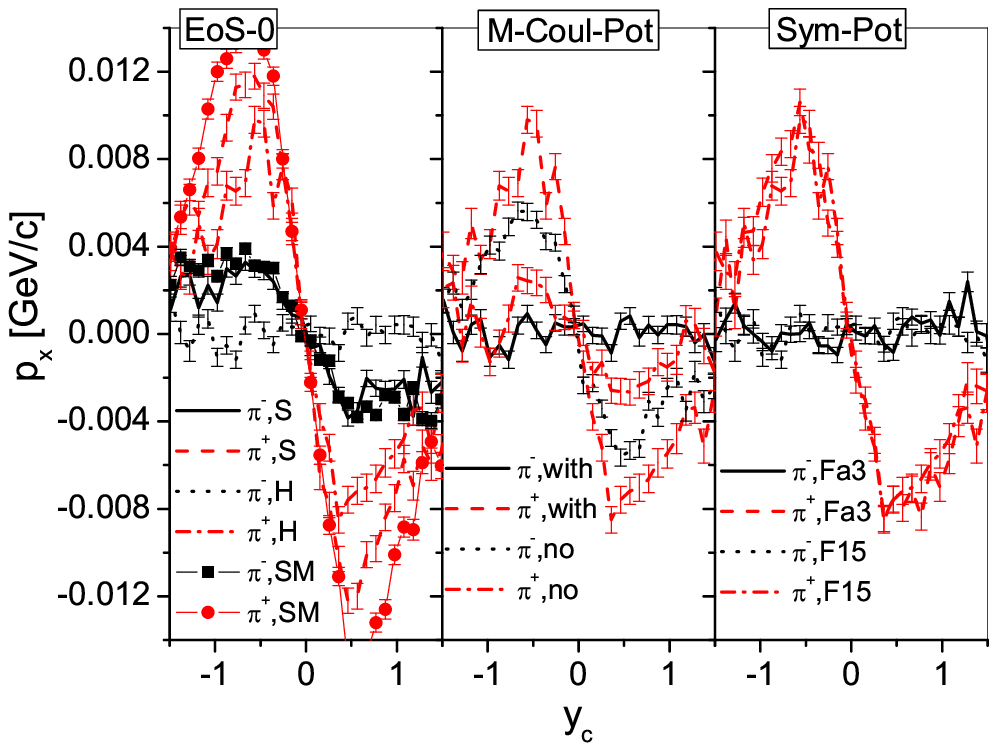}
\caption{The directed transverse flow distributions of the $\pi^-$
and $\pi^+$ mesons for various isospin-independent EoS (S-, H,
and SM-EoS, left plot), with or without mesonic Coulomb potential
(middle plot), and with the F15 and Fa3 symmetry potentials (right
plot) (see text). The reaction $^{208}{\rm Pb}+^{208}{\rm Pb}$ at $E_b=0.8A$
GeV and $b=7-9$ fm is chosen.}
\label{fig7}
\end{figure}

Now let us turn to discuss collective flow observables. Fig.\ \ref{fig7} shows the in-plane directed transverse flow of
charged pions as a function of rapidity. It is well
known that the directed flow of particles is particularly sensitive to the
stiffness of the EoS, which is also shown in the left plot of
Fig.\ \ref{fig7}. The weak anti-flow of charged pions for semi-peripheral collisions is well reproduced, similar to the
calculations and the experiments in Refs.
\cite{Bass95,LiB96,Kin97} for semi-peripheral Au+Au collisions at
different beam energies. Among the three types of the EoS studied
here, the anti-flow effect of $\pi^-$ and $\pi^+$ calculated with the
SM-Eos is the strongest, followed by the S-EoS \cite{Bass95}. The anti-flow effect of $\pi^+$ is much stronger
than that of $\pi^-$. If the mesonic Coulomb potential is neglected, the
absolute value of the antiflow parameter of $\pi^-$ is even higher
than that of $\pi^+$ as shown in the middle of Fig.\ \ref{fig7},
which is in disagreement with experimental
measurements \cite{Kin97}. In the right plot of Fig.\ \ref{fig7},
we find that the effect of the density dependence of the symmetry
potential on the charged pion flow is very weak, which means that
the in-plane directed flow of charged pions can provide more
accurate information on the isospin-independent part of the EoS
rather than on the isospin dependent part.

\begin{figure}
\includegraphics[angle=0,width=0.8\textwidth]{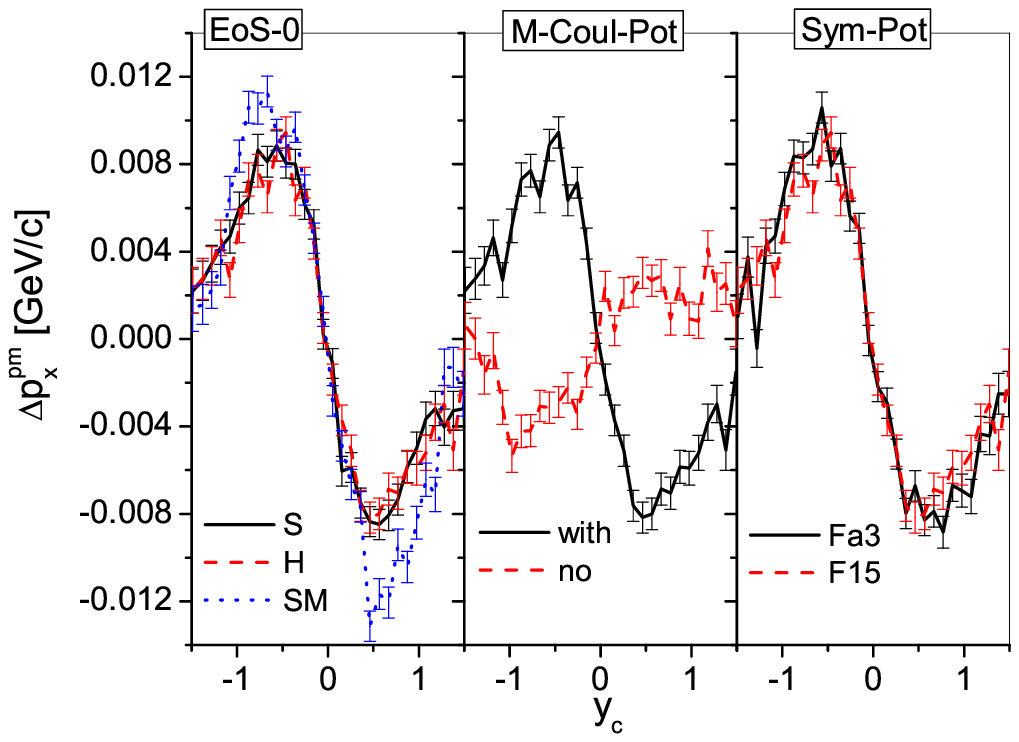}
\caption{The rapidity distributions of the $\pi^+ - \pi^-$ transverse flow difference ($\Delta p_x^{\rm
pm}=p_x^{\pi^+}-p_x^{\pi^-}$) for various isospin-independent EoS (S-, H,
and SM-EoS, left plot), with or without mesonic Coulomb potential
(middle plot), and with the F15 and Fa3 symmetry potentials (right
plot) (see text). The reaction $^{208}{\rm Pb}+^{208}{\rm Pb}$ at $E_b=0.8A$
GeV and $b=7-9$ fm is chosen.} \label{fig8}
\end{figure}

In Fig.\ \ref{fig8} we show the rapidity distributions of the
$\pi^+$-$\pi^-$ transverse flow difference $\Delta p_x^{\rm pm}$
defined by $\Delta p_x^{\rm pm}=p_x^{\pi^+}-p_x^{\pi^-}$. $p_x^{\rm pm}$ is similar when calculated with S-EoS, H-EoS as
well as SM-EoS at mid-rapidity, but it turns out to be different
between with the momentum dependence (SM-EoS) and without the
momentum dependence (S-EoS and H-EoS) in the
projectile-target rapidity region.  Again, from the right plot
we find that $\Delta p_x^{\rm pm}$ depends only weakly on the symmetry potential. Thus, we conclude that the $\pi^+$-$\pi^-$
transverse flow difference $\Delta p_x^{\rm pm}$ can be adopted to probe the
momentum dependence of the EoS.

\begin{figure}
\includegraphics[angle=0,width=0.8\textwidth]{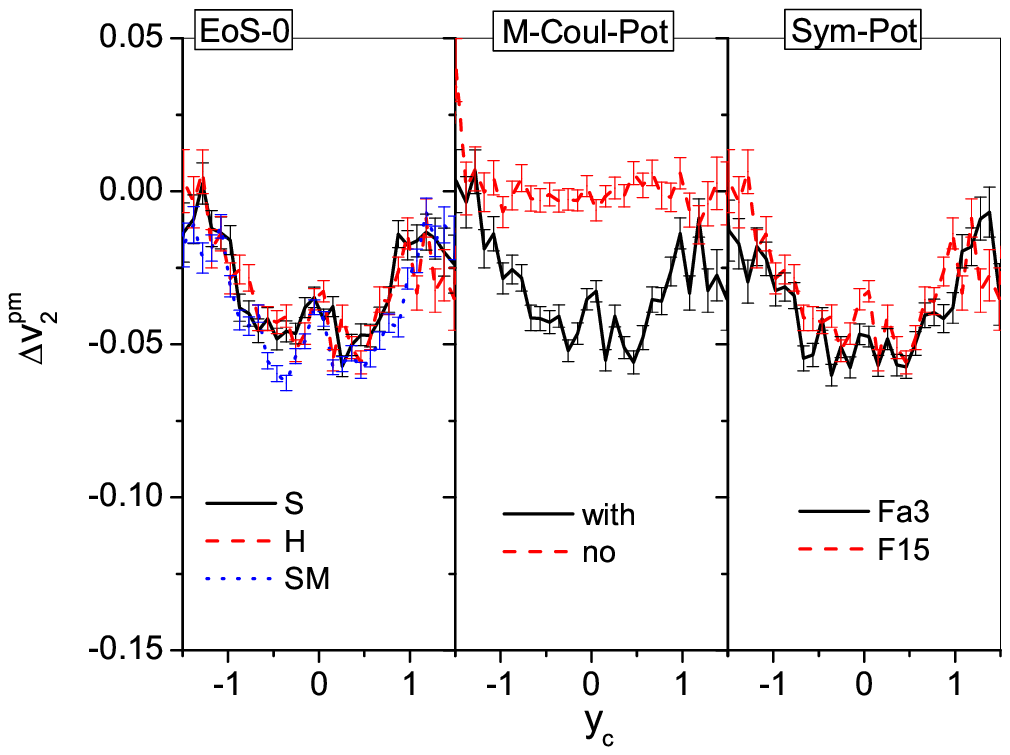}
\caption{Rapidity distributions of the $\pi^+ - \pi^-$ elliptic
flow difference ($\Delta v_2^{\rm pm}=v_2^{\pi^+}-v_2^{\pi^-}$) for various isospin-independent EoS (S-, H,
and SM-EoS, left plot), with or without mesonic Coulomb potential
(middle plot), and with the F15 and Fa3 symmetry potentials (right
plot) (see text). The reaction $^{208}{\rm Pb}+^{208}{\rm Pb}$ at $E_b=0.8A$
GeV and $b=7-9$ fm is chosen. }
\label{fig9}
\end{figure}

Fig.\ \ref{fig9} shows the influence of different parts in the EoS
on the $\pi^+$-$\pi^-$ elliptic flow difference as a function of
rapidity, which is defined as $\Delta v_2^{\rm
pm}=v_2^{\pi^+}-v_2^{\pi^-}$. Here
$v_2=\langle (\frac{p_x}{p_t})^2-(\frac{p_y}{p_t})^2 \rangle$ and $v_2^{\pi^+}$
is the elliptic flow for $\pi^+$, $v_2^{\pi^-}$ is the elliptic
flow for $\pi^-$. First, we see that $\Delta v_2^{\rm pm}$ value is
always negative when the mesonic Coulomb interaction
is taken into account. The
influence of the mesonic Coulomb interaction on the $\pi^+$-$\pi^-$
elliptic flow difference is shown in the middle plot of Fig.\ \ref{fig9}. $\Delta v_2^{\rm
pm}$ is almost zero when the mesonic Coulomb interaction is switched off.
When the mesonic Coulomb interaction is switched on, the out-of-plane elliptic flow
becomes larger for $\pi^+$ than for $\pi^-$ mesons
which leads to a negatively elliptic flow difference. The left
and right plots of Fig.\ \ref{fig9} show that $\Delta v_2^{\rm pm}$
is insensitive to the uncertainties of both the isospin-independent
and -dependent parts of the EoS. The effect of the momentum dependence
of the EoS has only little effect on $\Delta v_2^{\rm pm}$. The weak
dependence of the rapidity distribution of $\Delta v_2^{\rm pm}$ on the
mean-field potentials might be useful for extracting exclusive information on medium corrections of the binary
cross sections.

\begin{figure}
\includegraphics[angle=0,width=0.8\textwidth]{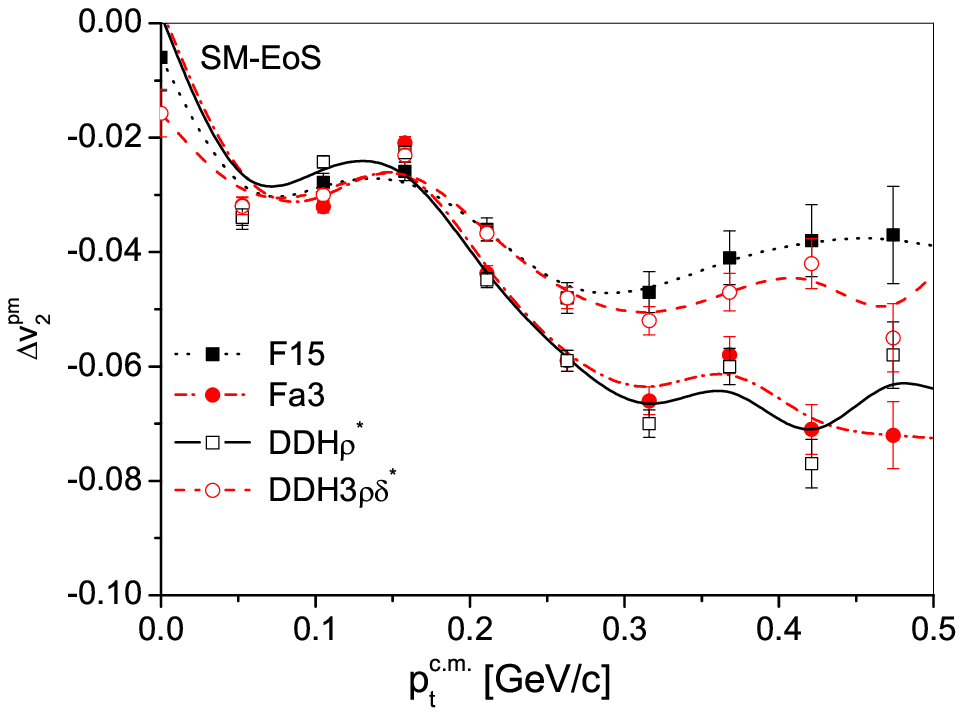}
\caption{The transverse momentum distribution of the $\Delta v_2^{\rm
pm}$ with different symmetry potentials. The SM-EoS is adopted for
the $^{208}{\rm Pb}+^{208}{\rm Pb}$ reaction at $E_b=0.8A$ GeV
and $b=7-9$ fm.} \label{fig10}
\end{figure}

From the above discussions on the pion flow (Figs.\
\ref{fig7}-\ref{fig9}), one might have the impression that the pion
flow is not sensitive to the density dependence of the symmetry
potential. Actually, this is not true, when we study the {\it
transverse momentum} dependence of $\Delta v_2^{\rm pm}$. In Fig.\
\ref{fig10} we illustrate $\Delta v_2^{\rm pm}$ of charged {\it
pions} as a function of transverse momentum $p_t^{c.m.}$. The SM-EoS is
adopted in the calculations. $\Delta v_2^{\rm pm}$ is negative
for all $p_t^{c.m.}$. The dependence of $\Delta v_2^{\rm pm}$ on the form of
the density dependence of the symmetry potential becomes very
pronounced at large transverse momenta $\sim 0.2-0.5$ GeV$/c$.
This behavior is very similar to the proton and neutron
elliptic flow difference $v_2^{\rm pn}$ \cite{Gre02,Gai04}. As it is known that pions
with high transverse momenta are mainly produced from the high
density region, Comparing Fig.\ \ref{fig10} with Fig.\
\ref{fig1}, we deduce that it reflects explicitly the density
dependence of the symmetry potential at densities $u>1$. At
densities $u>2.6$ (Fig.\ \ref{fig1}) the symmetry
potential Fa3 is smaller than DDH$\rho^*$ which is also
reflected in Fig.\ \ref{fig10} at the transverse momentum $\sim
0.5$ GeV$/c$.  At higher transverse momenta, although more than
1.6 million events were calculated, the statistical errors
are quite large and the results are not shown here.

\section{Summary and Outlook}
In summary, based on the UrQMD model (version 1.3) we have studied
the influence of different parts of the EoS such as the
isospin-independent, the Coulomb interaction (the Coulomb interaction between meson and meson (baryon) is also considered) and the symmetry
energy on a variety of observables related to
pion production in intermediate energy HICs. We first have studied the
isospin-symmetric central $^{40}{\rm Ca}+^{40}{\rm Ca}$ collisions. It shows
that the Coulomb interaction plays an important role in the reaction
dynamics and influences the rapidity and transverse momentum
distributions of charged pions considerably. Due to the
effect of the Coulomb potential, the $\pi^-/\pi^+$ ratio deviates from
unity and shows the dependence on the symmetry potential even for
isospin-symmetric systems. Our study shows that the rapidity
distributions of the pions and the $\pi^-/\pi^+$ ratio are
strongly influenced by the uncertainty of the EoS, while positively charged pions from neutron-rich HICs are
much less influenced by the density dependence of the symmetry
potential.

We have also studied the pion directed and
elliptic flow (difference) as functions of rapidity and transverse momentum
for the neutron-rich system $^{208}{\rm Pb}+^{208}{\rm Pb}$. We find that the
mesonic Coulomb interaction plays an important role in reproducing the proper
pion flow. The omission of mesonic Coulomb interaction leads to a wrong behavior of
the rapidity distribution of the $\pi^{-}/\pi^{+}$ ratio and the
charged pion directed and elliptic flow. We have also found that the
directed flow is sensitive to the isospin-independent EoS but not
sensitive to the various forms of the density dependence of the
symmetry potential. The rapidity distribution of the transverse flow difference $\Delta p_x^{\rm pm}$ of charged pions shows the
sensitivity only to the momentum dependence of the EoS. The
rapidity distribution of the elliptic flow difference $\Delta v_2^{\rm
pm}$ of charged pions is insensitive to the mean field
potentials. However, the transverse momentum distribution of $\Delta v_2^{\rm pm}$ of charged pions (at transverse momenta $\sim
0.2-0.5$ GeV$/c$) becomes more sensitive to the form of the
density dependence of the symmetry potential. Thus, we can use the
extensive comparison of multiple observables between the
calculations and the intermediate-energy experimental data to
extract comprehensive information on the EoS.

\section*{Acknowledgments}
Q. Li thanks the Alexander von Humboldt-Stiftung for a fellowship.
This work is partly supported by the National Natural Science
Foundation of China under Grant No.\ 10235030 and the Major State
Basic Research Development Program of China under Contract No.
G20000774, as well as by GSI, BMBF, DFG, and Volkswagenstiftung.

\end{document}